\documentclass[amsmath,amssymb,reqno,tbtags,psamsfonts,10pt,a4paper,twocolumn,showpacs, prl]{revtex4-1}
\usepackage[english]{babel}
\usepackage{stmaryrd}
\usepackage{color,graphicx}
\usepackage{upgreek}
\usepackage{dcolumn}
\usepackage{bm}
\usepackage[per-mode=symbol]{siunitx}

\usepackage{array}

\usepackage{times}
\usepackage{hyperref} 
\usepackage[capitalise]{cleveref}

\newcommand\eps{\ensuremath{\varepsilon}}

\newcommand{\av}[1]{{\ensuremath{\left\langle #1 \right\rangle}}}

\newcommand{\rss}{\ensuremath{r^\text{ss}_{bf}}}
\newcommand{\VSS}{\ensuremath{v^\text{ss}}}

\newcommand{\dd}{\mathrm{d}}

\definecolor{mygray}{gray}{0.4}
\newcommand{\TODO}[1]{}
\renewcommand{\TODO}[1]{\textcolor{mygray}{\textit{(#1)}}}

\begin{document}

\title[]{Collective motion of cells crawling on a substrate: roles of cell shape and contact inhibition}
\bigskip
\author{Simon K. Schnyder}
\email{skschnyder@gmail.com}
\affiliation{Department of Chemical Engineering, Kyoto University, Kyoto 615-8510, Japan}
\author{Yuki Tanaka}
\affiliation{Department of Chemical Engineering, Kyoto University, Kyoto 615-8510, Japan}
\author{John J. Molina}
\affiliation{Department of Chemical Engineering, Kyoto University, Kyoto 615-8510, Japan}
\author{Ryoichi Yamamoto}
\email{ryoichi@cheme.kyoto-u.ac.jp}
\affiliation{Department of Chemical Engineering, Kyoto University, Kyoto 615-8510, Japan}

\date{\today}

\begin{abstract}
Contact inhibition plays a crucial role in the motility of cells, the process of wound healing, and the formation of tumors. By mimicking the mechanical motion of cells crawling on a substrate using a pseudopod, we constructed a minimal model for migrating cells which gives rise to contact inhibition of locomotion (CIL) naturally. The model cell consists of two disks, one in the front (a pseudopod) and the other one in the back (cell body), connected by a finitely extensible spring. Despite the simplicity of the model, the cells' collective behavior is highly nontrivial, depending on the shape of cells and whether CIL is enabled or not. Cells with a small front disk (i.e. a narrow pseudopod) form immobile colonies. In contrast, cells with a large front disk (i.e. such as a lamellipodium) exhibit coherent migration without any explicit alignment mechanism being present in the model. This suggests that crawling cells often exhibit broad fronts because it helps them align. Upon increasing the density, the cells develop density waves which propagate against the direction of cell migration and finally arrest at higher densities. 
\end{abstract}

\pacs{
87.17.Jj, 
87.23.Cc, 
87.18.Fx, 
05.10.−a 
}
 
\maketitle

Directional collective motion of cells is of fundamental importance for embryogenesis, wound healing and tumor invasion \cite{Montell2008,Friedl2009,Rørth2009,Mehes2014, Mayor2016}. Cells move in clusters, strands or sheets to cover empty area~\cite{Kim2013}, to grow or invade tissues. How the cells coordinate and control their motion, is the subject of ongoing research. 
At the level of a single cell, it is well established that its motion is intricately linked to its shape. 
The shape of crawling cells is highly variable, depending on the type of cell, the substrate, as well as a result of the migration process itself \cite{Maeda2008, Keren2008, Ziebert2012,Ohta2015}. 
When a cell starts moving, its shape breaks symmetry \cite{Keren2008}, whereas circular cells typically cannot move. While there is evidence that shape has a strong influence on scattering and can lead to clustering and collective directed motion of swimmers \cite{Kantsler2013, Wensink2014}, less is known about the role of cell shape in organizing collective crawling.
It has been shown in simulations that inelastic collisions between crawling cells, e.g. due to deformation, can lead to coherent migration \cite{Grossman2008,Coburn2013,Ohta2014,Lober2015}, suggesting the importance of deformability for collective behavior.
When crawling cells come into contact, it inhibits their protrusions, which tends to change their shape and reorient them \cite{Abercrombie1979, Carmona-Fontaine2008}. It was shown that this effect, called contact inhibition of locomotion (CIL), enables cells to follow chemical gradients more effectively by aligning them \cite{Theveneau2010, Carmona-Fontaine2011}. In growing colonies, CIL leads to a slowing down of the motility of individual cells when the density of their environment crosses a certain threshold~\cite{Puliafito2012}. Thus, CIL is believed to play a crucial role in the control of collective tissue migration~\cite{Mayor2010,Theveneau2010, Theveneau2010a,Coburn2013}, tissue growth~\cite{Puliafito2012,Zimmermann2016}, morphogenesis, wound healing and in tumors~\cite{Li2014}.

Clearly, CIL, cell shape and deformability are linked \cite{Coburn2013}. So, we built a minimal, mechanical model of cells crawling on a substrate, aiming to isolate behavior purely caused by the interplay of contact inhibition and deformable shape, while neglecting properties such as cell-cell adhesion or chemotaxis. The simplicity of our model enables us to simulate considerably larger systems as compared to more complex models, which minimizes finite size effects.
The model is based on the accepted picture for a cell crawling on a surface  \cite{Lauffenburger1996, Ridley2003, Ananthakrishnan2007}: Before it begins migrating, the cell polarizes, i.e. front and back become distinguishable. 
Then the cell extends protrusions such as a pseudopod, driven forward by actin polymerization. The protrusions adhere to the substrate with adhesion sites, over which the cell exerts traction forces. Adhesion sites at the back of the cell are released and pulled in as the actin cytoskeleton depolymerizes.
Here, cells are represented by two disks, connected by a finitely extensible string. 
The cell migrates by expanding the spring, with the front disk exerting a motility force on the substrate.
We speculate on a mechanism for contact inhibition where the cell motility is proportional to the extension of the cell, motivated by the observation that cell speed depends on the extension of pseudopods \cite{Vedel2013}.
An alternative motility term where the force is always constant was used for comparison.

From a minimal model, quantitative agreement with experiments cannot be expected, but we find qualitative agreement with a wide range of properties of crawling cells.
As in cell colonies such as MDCK cells, we find that the average cell speed decreases strongly with cell density, an effect that vanishes when we switch contact inhibition off. The cell speed distributions are  similar to those of fibroblasts.
Further, we find a dynamic phase transition as a function of cell shape: 
When the front is larger than the back -- typical for many migrating cells such as keratocytes or fibroblasts \cite{Mogilner2009} -- the cells exhibit coherent migration, even though there is no explicit alignment mechanism included in the model. This suggests that the broad front often observed in crawling cells helps them achieve coherent motion.
When contact inhibition is switched off, we find weakened alignment, pointing to the relevance of CIL in the collective migration of cells.
The transition from disorder to order which occurs for keratocytes when their density is increased~\cite{Szabo2006}, arises in the model when cell noise is
included. Finally, before arresting at high density, the system
exhibits strong density and velocity fluctuations, where dense regions of arrest travel against the average direction of motion. This phenomenon is also seen as spontaneously arising traffic jams in traffic flow. Similar waves have been recently observed~\cite{Serra-Picamal2012} and thus our results could link those to contact inhibition.

\begin{figure}
  \centering
    \includegraphics[width=0.98\columnwidth]{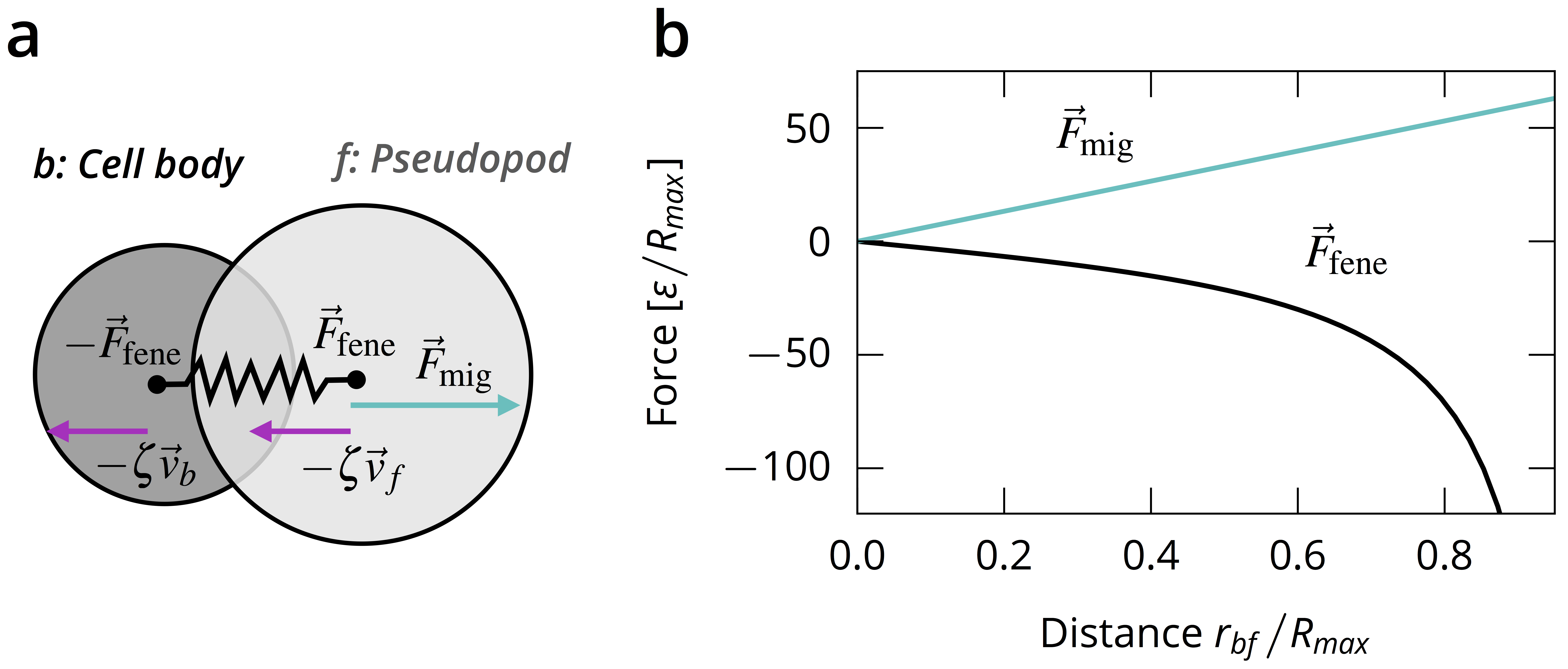}
  \caption{
	(a)	Schematic of the cell model.
	(b) Forces acting on the two disks being at distance
 $r_{bf}=|\vec r_{bf}|$ (\cref{eq:Fmig}).
}
  \label{fig:CellMigrationCycle}
  \label{fig:forces}
\end{figure}

\paragraph{Simulation}

Each cell consists of a cell body and a pseudopod, modelled as disks
with indices \emph{b} (back) and \emph{f} (front), respectively, see
\cref{fig:CellMigrationCycle}(a). 
The disks experience a drag force with the substrate $-\zeta_i \vec v_i$
with friction coefficient $\zeta_i$ and $\vec v_i$ being the velocity of
disk $i\in f,b$. Assuming that substrate friction is large compared to both
cell-cell friction and intracellular friction, we neglect the latter
two and set $\zeta = \zeta_b = \zeta_f$ for simplicity.
The two disks are connected by a finitely extensible nonlinear elastic
(FENE) spring \cite{Jin2007}, and the migration force $F_\text{mig}$ is applied only to the front disk as a linear function of the separation of the two disks
$\vec r_{bf}=\vec r_{f}-\vec r_{b}$, {\it i.e.},
\begin{align}
	\vec F_\text{fene}(\vec r_{bf}) = -\frac{\kappa \vec r_{bf}}{1-(r_{bf}/R_\text{max})^2},
	\label{eq:Fmig}
	\ \ \vec F_\text{mig} (\vec r_{bf}) = m \vec r_{bf}
\end{align}
with adjustable parameters $\kappa$, $m$, see
\cref{fig:forces}(b). $R_\text{max}$ as the maximum separation of the disks sets the characteristic length scale. 
This model can be derived from a more complex crawling model
(see supplemental materials), where cell locomotion
is achieved solely from repetitions of extensional and contractional motion of the cell using its cytoskeleton and adhesion to the substrate. 
The cell is only motile when its disks have some separation, $r_{bf}>0$,
and thus when its shape deviates from a circle. 
Such coupling of motility and deformation is typical in crawling cells \cite{Nelson2009}. 
The migration term $F_\text{mig}$ models contact inhibition, as cells compressed due to contact with their neighbors exert a lower motility force.
Disks of different cells interact via the short-ranged, purely repulsive Weeks-Chandler-Andersen potential~\cite{Weeks1971}, since interactions occur mainly via direct contact. All back disks have diameter $\sigma_b$, all front disks have diameter $\sigma_f$. To allow for different cell shapes, $\sigma_b$ and $\sigma_f$ can be different. The energy scale is set by $\eps$. (For details, see supplementary materials.)

For each of the cells we now have two coupled equations of motion, assuming overdamped dynamics,
\begin{gather}
	\begin{aligned}
		\label{eq:equationsOfMotion}
		\frac{\dd}{\dd t} \vec r_b &= \frac{1}{\zeta} \left(-\vec F_\text{fene}(\vec r_{bf}) + \sum_\text{neigh.} \vec F_\text{WCA} \right)\\
		\frac{\dd}{\dd t} \vec r_f &= \frac{1}{\zeta} \left(\vec F_\text{fene} (\vec r_{bf}) + \vec F_\text{mig}(\vec r_{bf})+ \sum_\text{neigh.} \vec F_\text{WCA} \right).
	\end{aligned}
\end{gather}
We chose $\kappa = 2.00\cdot 10^{4}\, \eps/R_\text{max}^2$ and $m = 4.14\cdot 10^{4}, \eps/R_\text{max}^2$, 
such that $m = 2.07\kappa$. 
Then, cells can enter a steady state of constant extension $\rss$ in which the forces acting on the cell balance,
\begin{gather*}
0= 
\frac{\dd}{\dd t} r_{bf} 
= \frac{1}{\zeta} \left(m \rss 
- \frac{2\kappa \rss}{1-(\rss/R_\text{max})^2}\right).
\end{gather*}
Thus, steady-state distance $r_{bf}^{ss}$ and the corresponding cell velocity $\VSS$ are given by
\begin{gather}
	\begin{aligned}
		\rss = R_\text{max}\sqrt{1 -2\kappa/m },\ \  
		\VSS = \rss m /(2\zeta).
	\end{aligned}
\end{gather}
With these parameters, the cells' length is of order $R_\text{max}$, see table SI.
The characteristic time scale of migration, 
$\tau_\text{mig} = R_\text{max}/\VSS = 2.06\cdot 10^{-5}\zeta R_\text{max}/\eps$,
is the time it takes for a solitary cell in the steady state to travel roughly its own length. 

For comparison, we use a version of the model without CIL. Replacing the migration force term $m r_{bf}$ with constant value $m \rss$ leaves $\rss$ and $\VSS$ unchanged, but leads to cells always exerting exactly the same migration force, regardless of whether the local environment allows for extension of the cell. This makes the system more similar to Vicsek-type models with constant speed \cite{Vicsek1995,Gregoire2004,Chate2008}. 

Cells are placed on random positions in a square simulation box of length $L$ with periodic boundary conditions at area fraction $\varphi = A N/L^{2}$, with the total number of cells $N$ and the area of a single cell in its steady state $A \approx 0.29 R_\text{max}^2$.
Configurations at large $\varphi$ were obtained from systems at lower area fraction and randomly letting individual cells divide into two new cells if there was enough space. To minimize finite size effects, we simulated systems with up to $10^5$ cells. 
We integrated the equations of motion until the steady-state was reached. 
All results are averaged over 10 independent runs.

\begin{figure}
\includegraphics[width = \columnwidth]{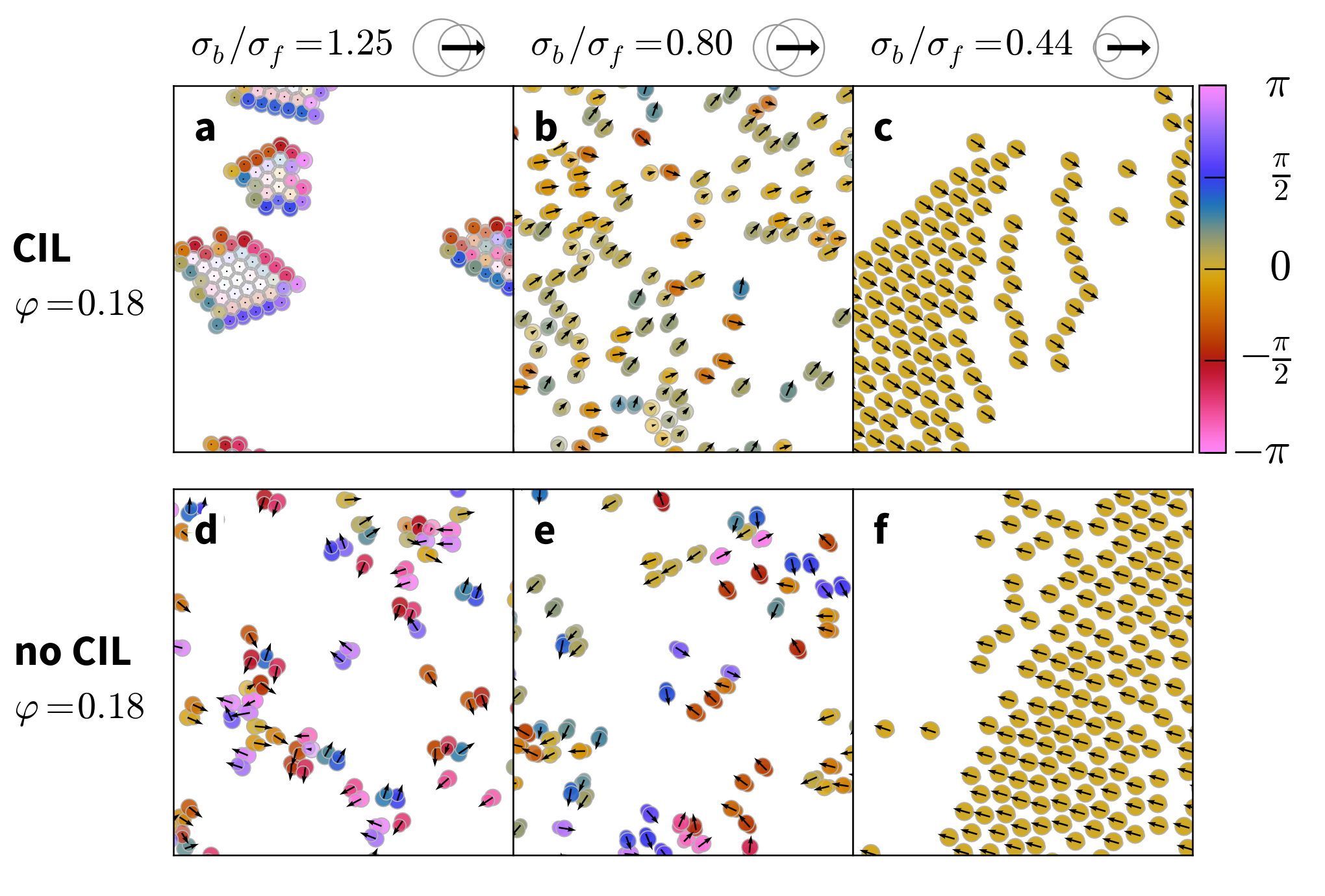}
  \caption{Snapshots of CIL and no-CIL cells for a range of cell shapes. Cell velocities are given as arrows and color. Hue indicates deviation from average direction, and slower cells are lighter in color. (For videos, see supplementary materials)}
  \label{fig:snapshotSmallMultiples}
\end{figure}

\paragraph{Results}

In order to investigate the influence of the cell shape on their
dynamics, we varied the diameters $\sigma_b$ and $\sigma_f$ of the cell
disks while keeping the area of the cell in the steady state constant. 
The model has no random component, which is a valid assumption when the dynamics are dominated by collisions \cite{Drescher2011,Wensink2012} -- a reasonable assumption at intermediate and high cell densities.
When the back disk is bigger than the front, $\sigma_b > \sigma_f$, the cells tend to form mostly immobile colonies, see \cref{fig:snapshotSmallMultiples}a).
When the front disk is larger than the back, $\sigma_b < \sigma_f$, the cells exhibit coherent migration, see \cref{fig:snapshotSmallMultiples}b). If the front is much larger than the back, the cells completely align and form dense, travelling bands, see \cref{fig:snapshotSmallMultiples}c).  This behavior is quite similar to that of migrating neural crest cells \cite{Carmona-Fontaine2008, Carmona-Fontaine2011} and occurs here without requiring cell-cell attraction.
 Uninhibited cells do not form colonies and exhibit weaker alignment at $\sigma_b/\sigma_f = 0.80$.

 \begin{figure}
   \centering
     \includegraphics[width=\columnwidth]{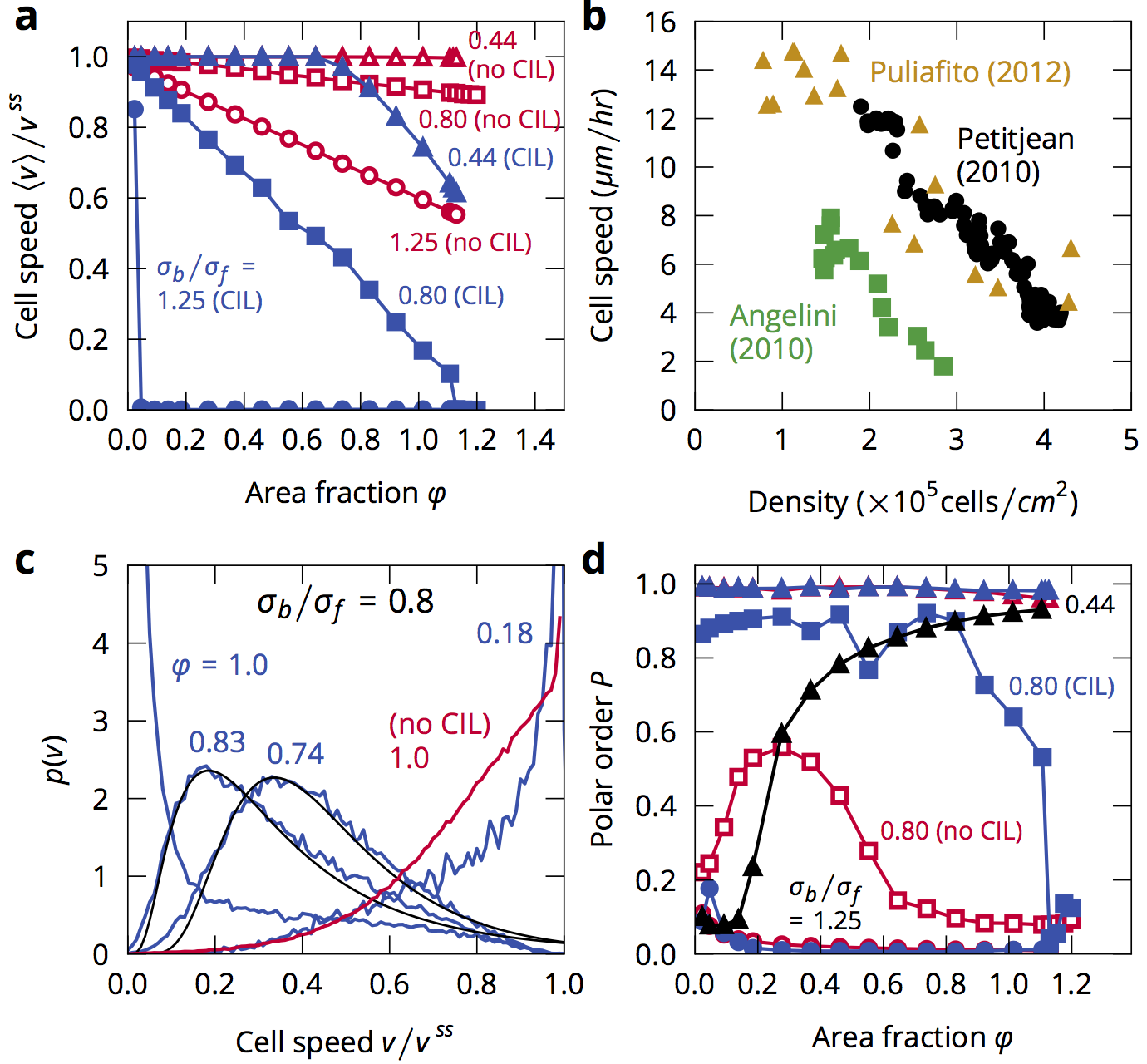}
   \caption{(a) Average cell speed, normalized by the steady state speed
  of a solitary cell $\VSS$, with (closed symbols) and without contact
  inhibition (open symbols) and a range of cell shapes. (b) Speed of
  MDCK cells in a growing cell colony (extracted from
  \cite{Puliafito2012}), and in confluent monolayers (extracted from
  \cite{Petitjean2010,Angelini2011}). (c) Speed histograms at
  $\sigma_b/\sigma_f = 0.80$. Black lines are fits to the data with a
  log-normal distribution. (d) Polar order $p$ for a range of cell
  shapes. Simulations with cell noise at $\sigma_b/\sigma_f = 0.44$ are shown in black.}
 	  \label{fig:petitjean_velocity_vs_density}
 	  \label{fig:velocityDistributions}
   \label{fig:comboplot}
   \label{fig:velocities2}
 \end{figure}

Migrating cells slow down strongly at high cell densities~\cite{Petitjean2010, Angelini2011, Puliafito2012,Doxzen2013}. To test this in the model, we measured the average cell speed while varying the area fraction of the cells at three shape anisotropies $\sigma_b/\sigma_f = 1.25$, $0.80$ (the inverse case), and $0.44$, see \cref{fig:velocities2}. For $\sigma_b/\sigma_f = 1.25$, the cell speed vanishes for all but the smallest density due to formation of jammed clusters. In the reverse case, $\sigma_b/\sigma_f = 0.80$, the contact inhibited cells crawl at maximum speed only at very small density, with the speed decreasing linearly over the whole density range. The cells fully arrest when they are close-packed, at $\varphi \approx 1.1$ (Area fractions can be larger than 1 because the disks are soft, and because the area fraction is defined with the cell's biggest possible area, in its steady state).
At $\sigma_b/\sigma_f = 0.44$, cells crawl at maximum speed up to $\varphi \approx 0.6$ where a slowing down occurs. 
In comparison, the uninhibited cells at corresponding anisotropies show a much weaker response to increasing density. 
The strong slowing down of the CIL cells with big fronts is qualitatively
comparable to the behavior of MDCK cells, see \cref{fig:comboplot}b). Even though MDCK cells are  
adhesive with a wide variability in cell area, while our cells are not
adhesive and vary their area only little, our model's
 coupling of cell extension and motility already gives rise to the known
relationship between cell density and speed. Note that
\citet{Garcia2015} recently suggested that the slowing down in MDCK
cells may not be driven by density primarily but by cell-cell and
cell-substrate adhesion. This is not captured in our model. 

The distribution of cell speeds in the aligning case, e.g. $\sigma_b/\sigma_f = 0.80$, see \cref{fig:velocityDistributions}c),
depends strongly on density as well. At low densities, most cells
move at maximum speed $v^{ss}$, while at high densities most cells are arrested. At intermediate densities, a distribution with a non-Gaussian tail arises. Such distributions are found in fibroblasts in monolayers \cite{Selmeczi2005,Nnetu2012,Vedel2013}.
Here, this distribution only arises in the CIL cells, pointing to the relevance of CIL in establishing typical collective crawling cell dynamics. A key difference to the experiments is that \citet{Vedel2013} report that the speed distribution does not depend on density which is impossible here because of the slowing down effect of our CIL mechanism.

Coherent motion of cells can be measured with the polar order $P$, an order parameter of collective alignment
\begin{align}
	\label{eq:orderParameter}
	P = \left|\av{\vec r_{bf}/|\vec r_{bf}|}\right|,
\end{align}
which evaluates to 1 for full alignment of the cells and to 0 for fully random or isotropic orientations, see \cref{fig:comboplot}d). When the CIL cells jam into clusters, e.g. at $\sigma_b/\sigma_f = 1.25$, the order parameter vanishes, as the  orientations of the cells mostly point towards their cluster's center. This leads to vanishing average speed. 
At $\sigma_b/\sigma_f = 0.80$, the cells are mostly aligned for most densities. Alignment weakens in the approach to full arrest. 
At $\sigma_b/\sigma_f = 0.44$ the cells are completely aligned at all densities. At densities where it is possible for the cells to be spaced far enough from each other so as not to interact (for roughly $\varphi \leq 0.65$) there are little to no collisions and the cells crawl at full speed. For $\sigma_b/\sigma_f = 0.80$, where alignment is not perfect, frequent collisions lower the average speed of the coherently moving cells. 
The non-CIL cells at $\sigma_b/\sigma_f = 1.25$ do not form clusters, instead moving disorderedly with vanishing polar order. At $\sigma_b/\sigma_f = 0.80$, the cells show some alignment, especially at intermediate densities, but are always more weakly aligned than corresponding CIL cells. At $\sigma_b/\sigma_f = 0.44$, the cells achieve near perfect orientational order at all densities, just as the CIL cells.

CIL cells with small fronts cluster because their motility tends to stay
pointed towards the other cells after a collision, which compresses and inhibits them. Similar jamming is also observed in active particle systems \cite{Buttinoni2013, Bechinger2016}.
Whereas the jamming of cells purely due to their shape has not been observed yet, it has been shown that crawling cells may form tissue-like clusters when placed on soft substrates but scatter apart on stiff substrates \cite{Guo2006}. Cells on stiff substrates are able to exert stronger traction forces than on soft substrates \cite{Wang2012}. Thus, even though we do not model substrate stiffness, the result that our CIL cells cluster while uninhibited cells (which on average exert stronger traction forces) do not, is then in rough qualitative agreement with the finding of \citet{Guo2006}.

Both CIL and no-CIL cells undergo a transition from disorder to coherent migration, driven by the shape asymmetry of the cells. 
While such a transition as function of cell shape has not been observed in experiments, it may explain why crawling cells often exhibit a broad front: \emph{It improves alignment.}
A similar transition occurs in the self-assembly of roughly triangular, stiff, active particles, but we find the transition \emph{reversed}: cells with a big front travel coherently here and cluster in \cite{Wensink2014}, while cells with a small front cluster here and travel coherently in \cite{Wensink2014}. The most notable difference between the models and likely the reason behind the reversal is that cells are highly deformable: they are easily compressed during collisions, which changes the collision dynamics strongly. 

Our model suggests that contact inhibition -- among other factors -- enhances the alignment of crawling cells, in agreement with results for neural crest cells \cite{Carmona-Fontaine2008, Carmona-Fontaine2011}.

Keratocytes exhibit a different transition from disordered to coherent
motion, driven by an increase in density \cite{Szabo2006}. At first,
this behavior is not found in our model, since the alignment of cells is
mostly independent of density. However, by including a random force
acting on the cell disks (see supplementary materials for details), this
transition occurs in the model as well, see the black line with triangles in
\cref{fig:comboplot}d) and Fig.~S2. In contrast to most models where the alignment mechanism
is included explicitly, alignment due to cell shape is
responsible here.

\begin{figure}   
	\includegraphics{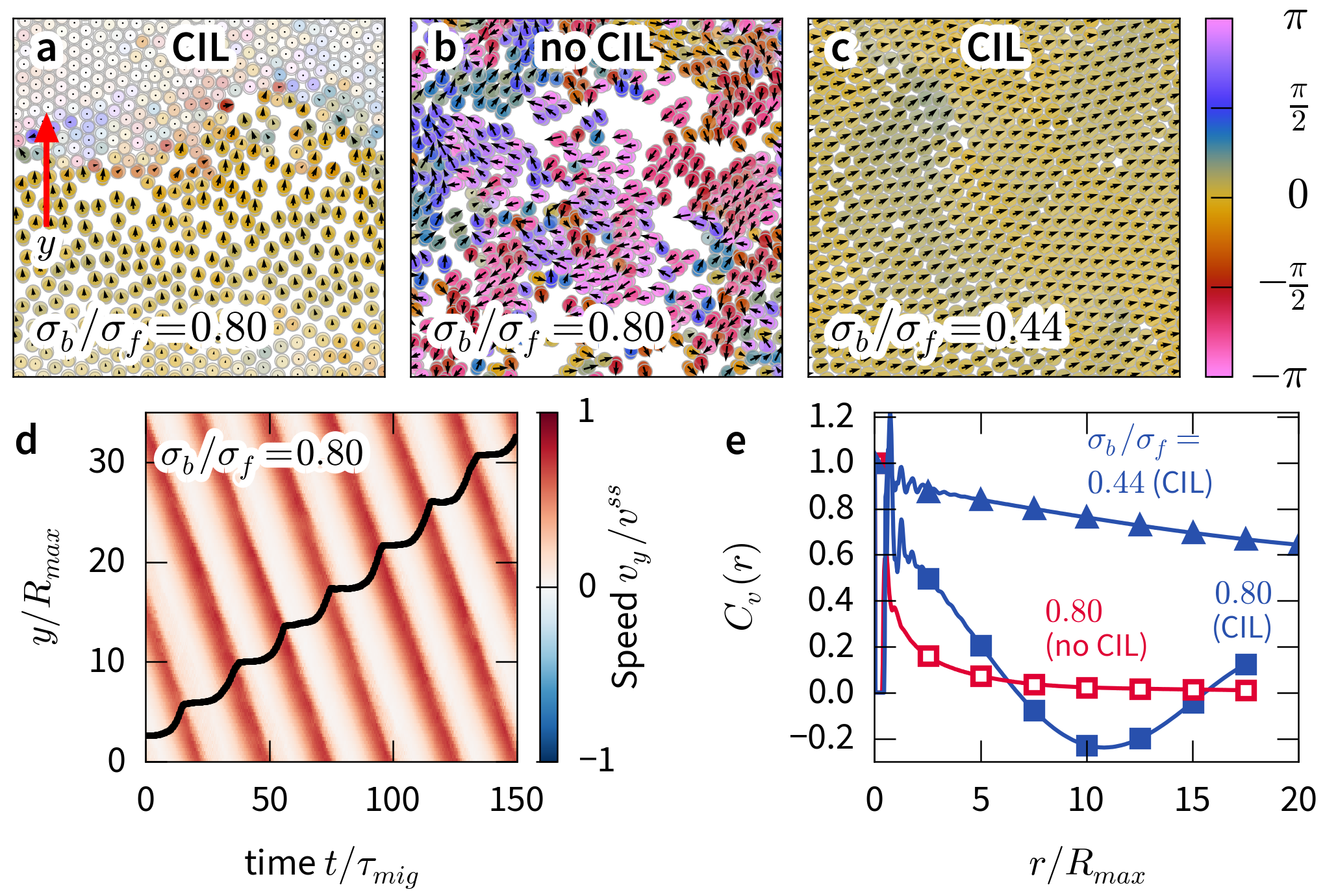}
  \caption{(a-c) Simulation snapshots of CIL and no-CIL cells at area fraction $\phi = 0.92$ with cell velocities shown as arrows. Hue indicates deviation from average direction, and slower cells are lighter in color. (For videos, see supplementary materials)
	(d) Kymograph of simulation shown in (a): Average of cell velocities in y-direction, see arrow in (a), along the y-direction as a function of time. Trajectory of one cell superimposed in black.
(e) Velocity correlation functions for simulations shown in (a-c).
	}
	\label{fig:trafficJam}
  \label{fig:kymograph}
\end{figure}

In the transition into arrest of cells with $\sigma_b/\sigma_f = 0.80$, a remarkable
feature develops: areas of dynamic arrest form and dissolve again. The
arrested areas grow in size with density until they
become system-spanning waves, see \cref{fig:trafficJam}a), suggesting a growing length scale. The
waves travel against the direction of motion of the cells, see
\cref{fig:kymograph}d), akin to traffic jams in models for car traffic
\cite{DeWijn2012,Nagel1992, Helbing2001}. 
In this state, the cell speed distribution shows a peak at or near 0 and a long tail, see data for $\varphi = 1.0$ in \cref{fig:comboplot}c).
The onset of system-spanning arrest waves roughly coincides with the decrease of the order parameter at $\varphi \approx 0.85$.
We don't observe such waves in the non-CIL systems, see \cref{fig:trafficJam}b), thus directly connecting the waves to CIL. 
Further, the waves only occur when contact inhibition is strong: The cells
with $\sigma_b/\sigma_f = 0.44$ -- whose slowing down at
high densities is weaker -- 
always travel coherently, \cref{fig:trafficJam}c). The reason for the weaker contact inhibition lies in their shape: When cells collide, they are compressed. If the size asymmetry is more extreme, the distance $r_{bf}$ in
the most compressed state tends to be bigger. Since the motility force
is proportional to $r_{bf}$, more asymmetric cells are less inhibited and their slowing down is less pronounced. This restraint in breaking can suppress jams in traffic models and thus explains the qualitative difference between the two systems. 
The correlation function of cell speed fluctuations 
$C_v(r) = \av{\Delta \vec v(0)\cdot \Delta \vec v(r)}/\av{\Delta \vec v(0)^2}$,
with $\Delta \vec v(r) = \vec v(r) - \av{\vec v}$, becomes negative on the length scale of the extent of the traffic jam, see \cref{fig:trafficJam}e). This is different from the quick decay in the disordered state of the uninhibited cells and the slow decay of the highly ordered state at $\sigma_b/\sigma_f = 0.44$.

Similar backwards traveling waves have been observed in expanding
monolayer sheets of MDCK cells~\cite{Serra-Picamal2012} but it remains open if
waves with full arrest can occur in crawling cells, e.~g. in ring geometries \cite{Li2014b,Doxzen2013}. 
Since our CIL mechanism links the velocity waves to corresponding density waves, it makes them distinct from heterogeneous velocity fields occurring without corresponding heterogeneous density \cite{Angelini2011}.

\paragraph{Summary}
In order to reveal universal dynamics of contact inhibited, deformable
cells, we modelled crawling cells on a substrate in a minimal,
mechanical model where cell motility was motivated by the internal dynamics of the cells. We assumed the motility force to be proportional to the extension of the cell, thus giving rise to contact inhibition of locomotion naturally. We find rich dynamic behavior in qualitative agreement with a variety of experiments, with multiple phase transitions as a function of cell shape, cell density and whether locomotion is inhibited or not.
Our results may explain why crawling cells often exhibit a broad front: It enhances alignment. Finally, we find density waves that propagate against the direction of cell motion. 

This model is a natural candidate to further investigate the dynamics of cellular tissues. Of particular interest would be the effect of contact inhibition and cell shape on tissue growth and wound closure, and the dynamics of malignant cells in mixtures of contact-inhibited and uninhibited cells.

\begin{acknowledgments}
We thank Estelle Gauquelin, J\"urgen Horbach, Benoit Ladoux, Norihiro Oyama, and Mitsusuke Tarama for helpful discussions. 
We acknowledge support by the Japan Society for the Promotion of Science (JSPS) KAKENHI Grant No. 26247069 and 26610131, and the supporting program for interaction-based initiative team studies (SPIRITS) of Kyoto University.
\end{acknowledgments}

\bibliographystyle{apsrev-mod}
\bibliography{CollectiveMotion}
\end{document}


\title[]{Collective motion of cells crawling on a substrate: roles of cell shape and contact inhibition}
\bigskip
\author{Simon K. Schnyder}
\email{skschnyder@gmail.com}
\affiliation{Department of Chemical Engineering, Kyoto University, Kyoto 615-8510, Japan}
\author{Yuki Tanaka}
\affiliation{Department of Chemical Engineering, Kyoto University, Kyoto 615-8510, Japan}
\author{John J. Molina}
\affiliation{Department of Chemical Engineering, Kyoto University, Kyoto 615-8510, Japan}
\author{Ryoichi Yamamoto}
\email{ryoichi@cheme.kyoto-u.ac.jp}
\affiliation{Department of Chemical Engineering, Kyoto University, Kyoto 615-8510, Japan}

\date{\today}

\widetext
\clearpage
\begin{center}
\textbf{\large Supplemental Materials:\\ Collective motion of cells crawling on a substrate: roles of cell shape and contact inhibition}
\end{center}
\setcounter{equation}{0}
\setcounter{figure}{0}
\setcounter{table}{0}
\setcounter{page}{1}
\makeatletter
\renewcommand{\theequation}{S\arabic{equation}}
\renewcommand{\thefigure}{S\arabic{figure}}
\renewcommand{\thetable}{S\Roman{figure}}

\section{Derivation of the model from explicit crawling motion}

\begin{figure}[h]
  \centering
    \includegraphics[width=0.8\columnwidth]{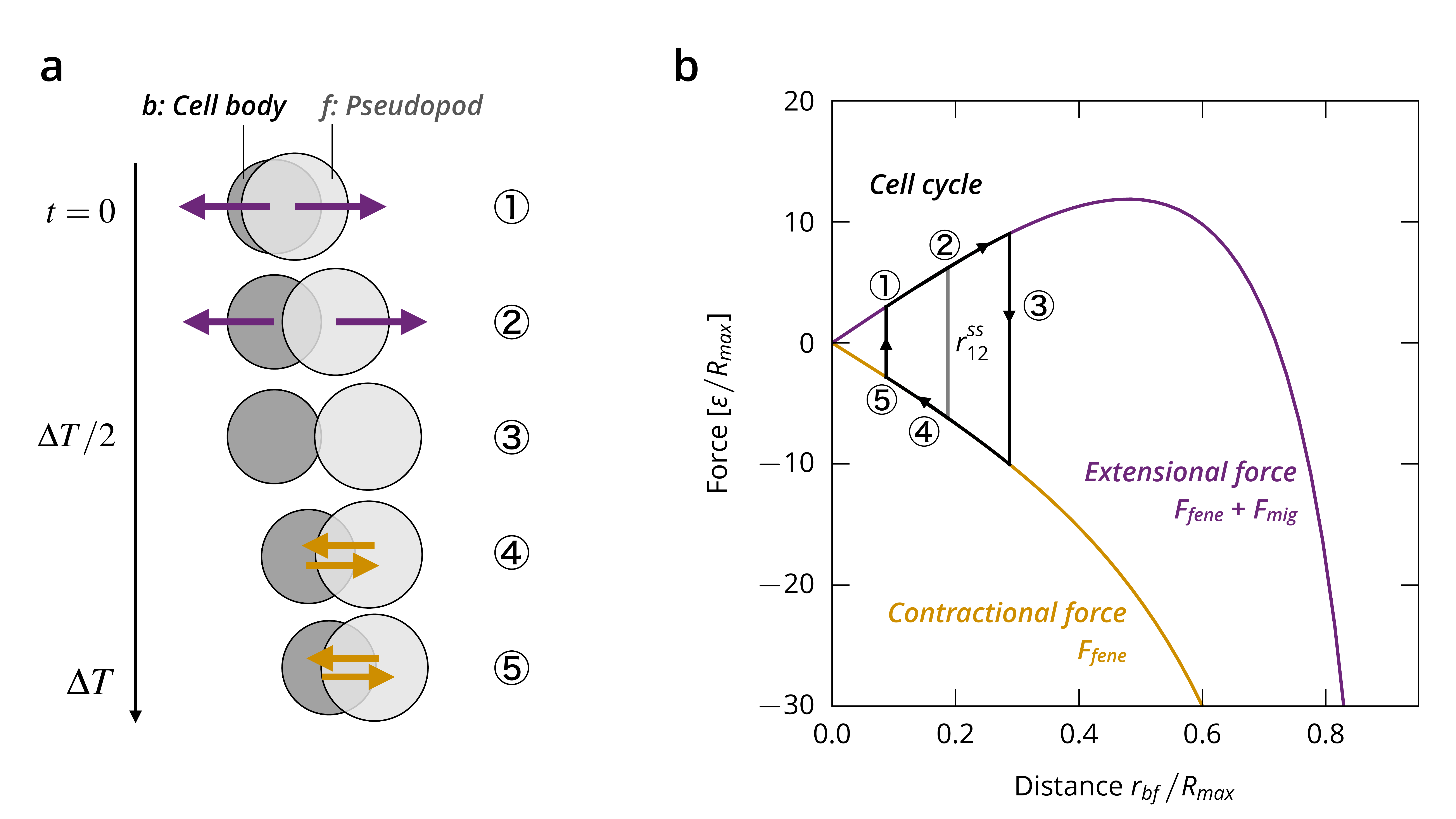}
  \caption{
	(a)	The cell migration cycle which underlies our coarse-grained mechanics.
	(b) The forces acting in a single cell on the two disks being at distance $r_{bf}$. The two-phase cell migration cycle of (a) is marked as a black path. In the limit of vanishing cycle period, the cell obtains constant extension $\rss$ where the forces exactly balance (marked by grey line).}
  \label{fig:CellMigrationCycle_smaller}
\end{figure}

The cell migration mechanism presented in this letter can be derived
from a coarse-graining of a two stage periodic crawling cycle that is
repeated cyclically with a time period $\Delta T$, see
\cref{fig:CellMigrationCycle_smaller}. 
In the first stage of the crawling motion $0 < t < \Delta T/2$, 
the pseudopod is pushed forward against friction with the substrate 
by an extensional force, $F_\text{fene}+F_\text{mig}$, 
which acts between the two disks of opposite sign, while 
the cell body adheres to the substrate with $\zeta_b=\infty$,
\begin{align}
\vec	v_b(t) =0,\ \ \  \vec v_f(t)=\frac{\vec F_\text{fene}+\vec F_\text{mig}}{\zeta_f}.
\end{align}
For times $\Delta T/2 < t < \Delta T$, 
the cell is in the second stage in which the pseudopod adheres to the
substrate $\zeta_f=\infty$ and the cell body is drawn in with a
contractional force $\vec F_\text{fene}$. 
\begin{align}
\vec	v_b(t) =\frac{-\vec F_\text{fene}}{\zeta_b},\ \ \  \vec v_f(t)=0.
\end{align}
The average velocities of back and front disks over the cycle $\Delta T$
are then given by
\begin{align}
\bar{\vec	v}_f &= \frac{1}{\zeta_f\Delta T}\int_{0}^{\Delta
 T/2}[\vec F_\text{fene}(\vec r_{bf}(t))+F_\text{mig}(\vec
 r_{bf}(t))] dt,\\
\bar{\vec	v}_b &= -\frac{1}{\zeta_b\Delta T}\int_{\Delta T/2}^{\Delta T}\vec F_\text{fene}(\vec r_{bf}(t))dt.
\end{align}

In the limit
$\Delta T \to 0$, $\vec r_{bf}(t)$ doesn't change during the cycle thus,
\begin{align}
\vec	v_f(t) &= \frac{1}{2\zeta_f}[\vec F_\text{fene}(\vec r_{bf}(t))+F_\text{mig}(\vec r_{bf}(t))],\\
\vec	v_b(t) &= -\frac{1}{2\zeta_b}\vec F_\text{fene}(\vec r_{bf}(t)).
\end{align}
Eq.~(2) is finally obtained by using $\zeta=2\zeta_b=2\zeta_f$ and adding
the cell-cell interaction force $F_\text{WCA}$.
The original alternated application of each force is replaced by the
application of both forces at the same time in eq.~(2). This can
be seen as the assumption that in a real cell, contraction, expansion,
as well as fixing and unfixing of body and pseudopod happen at the same
time or in close succession. 

\section{Cell-cell interaction}

The interaction between disks of different cells is modelled with the short-ranged Weeks-Chandler-Andersen potential~\cite{Weeks1971}, since interactions occur mainly via direct contact. All back disks have diameter $\sigma_b$, all front disks have diameter $\sigma_f$. To allow for different cell shapes, $\sigma_b$ and $\sigma_f$ can be different. For the interaction of a pair of disks $\alpha$ and $\beta$ ($\alpha, \beta \in [b,f]$) of two different cells at distance $\vec r$, the interaction diameter is given by $\sigma_{\alpha\beta} = (\sigma_\alpha + \sigma_\beta)/2$, the energy scale is given by $\eps$, and the force by
\begin{align}
	\vec F_\text{LJ}(r) = 
	\begin{cases}
		-24\eps \left[2\left(\frac{\sigma_{\alpha\beta}}{r}\right)^{12} - \left(\frac{\sigma_{\alpha\beta}}{r}\right)^6\right]\vec r/r^2, & r < r\cut, \\
		0, & r \geq r\cut.
	\end{cases}
\end{align}
The cutoff at $r\cut = 2^{1/6} \sigma_{\alpha\beta}$ makes the
present inter-cellular forces purely repulsive, but inclusion of attractive
terms modelling inter-cellular adhesion would be straightforward.

\section{Cell area}

The area $A$ of a cell with a back disk of diameter $\sigma_b$, a front disk of diameter $\sigma_f$ and a distance $r_{bf}$ between the particles is given by
\begin{align}
	A &= A_1 + A_2 - \text{overlap} \nonumber\\
	& = \frac{\pi}{4} (\sigma_b^2 + \sigma_f^2) 
	 - \frac{\sigma_b^2}{4} \cdot \cos^{-1} \left(\frac{4r_{bf}^2 + \sigma_b^2 - \sigma_f^2}{8 r_{bf}\sigma_b}\right)
	 - \frac{\sigma_f^2}{4} \cdot \cos^{-1} \left(\frac{4r_{bf}^2 - \sigma_b^2 + \sigma_f^2}{8 r_{bf}\sigma_f}\right) \nonumber\\
	&+ \frac{1}{8} \sqrt{(-2r_{bf} + \sigma_b + \sigma_f)(2r_{bf} + \sigma_b + \sigma_f)} 
	\cdot \sqrt{(2r_{bf} + \sigma_b - \sigma_f)(2r_{bf} - \sigma_b + \sigma_f)}
\end{align}

We fix the diameters of the disks such that at the steady-state distance
$\rss$, the area of the cells is constant, $A = 0.29 R^2_\text{max}$,
regardless of the shape, i.e. the shape anisotropy
$\sigma_b/\sigma_f$. For the three shapes, the cell sizes are given in
Table SI. The length of the cells thus is then always of order $R_\text{max}$.

\begin{table}[htb!]
\begin{tabular}{llll}
	\toprule
$\sigma_b/\sigma_f$ & $\sigma_b$ & $\sigma_f$ & Cell length\\
	\toprule
 1.25 & 0.55$R_\text{max}$	& 0.44$R_\text{max}$	& 0.68$R_\text{max}$ \\
 0.80 & 0.44$R_\text{max}$	& 0.55$R_\text{max}$	& 0.68$R_\text{max}$ \\
 0.44 & 0.27$R_\text{max}$	& 0.60$R_\text{max}$	& 0.62$R_\text{max}$ \\
\end{tabular}
	\label{table:sizes}
\caption{Size parameters for the cells. Cell length is given as the length of the cell in its steady state $(\sigma_b + \sigma_f)/2 + \rss$.}
\end{table}

\section{Cell noise}

The dynamics of individual cells are often observed to fluctuate in time and space.
These apparently random differences and fluctuations can have important biological and medical consequences.
We implemented cell noise to allow for comparison to Keratocytes. The noise, which is applied to all cell disks, is given by the force 
\begin{align}
	\vec F_\text{noise} = \sqrt{2d}\, \vec\xi(t)
\end{align}
with both components of $\vec \xi(t) = (\xi_x(t), \xi_y(t))$ being normally
distributed random variables obeying
$\av{\xi_i(t)} = 0$ and
$\av{\xi_i(t)\xi_j(t')} = \delta_{ij} \delta(t - t')$ with Kronecker-$\delta$ $\delta_{ij}$ and $\delta$-function $\delta(t - t')$.
Since the simulation is performed with a finite time step $\Delta t$, the force per timestep is 
\begin{align}
	\vec F_\text{noise} = \sqrt{2d/\Delta t}\, \vec\xi(t)
\end{align}
The timestep is $1.9\cdot 10^{-4}\tau_\text{mig}$.
The black line in Fig.~3d) was calculated for $d = 8.0\cdot 10^{-4} R_\text{max}\VSS$.

\begin{figure}   
  \centering
    \includegraphics[width=0.8\columnwidth]{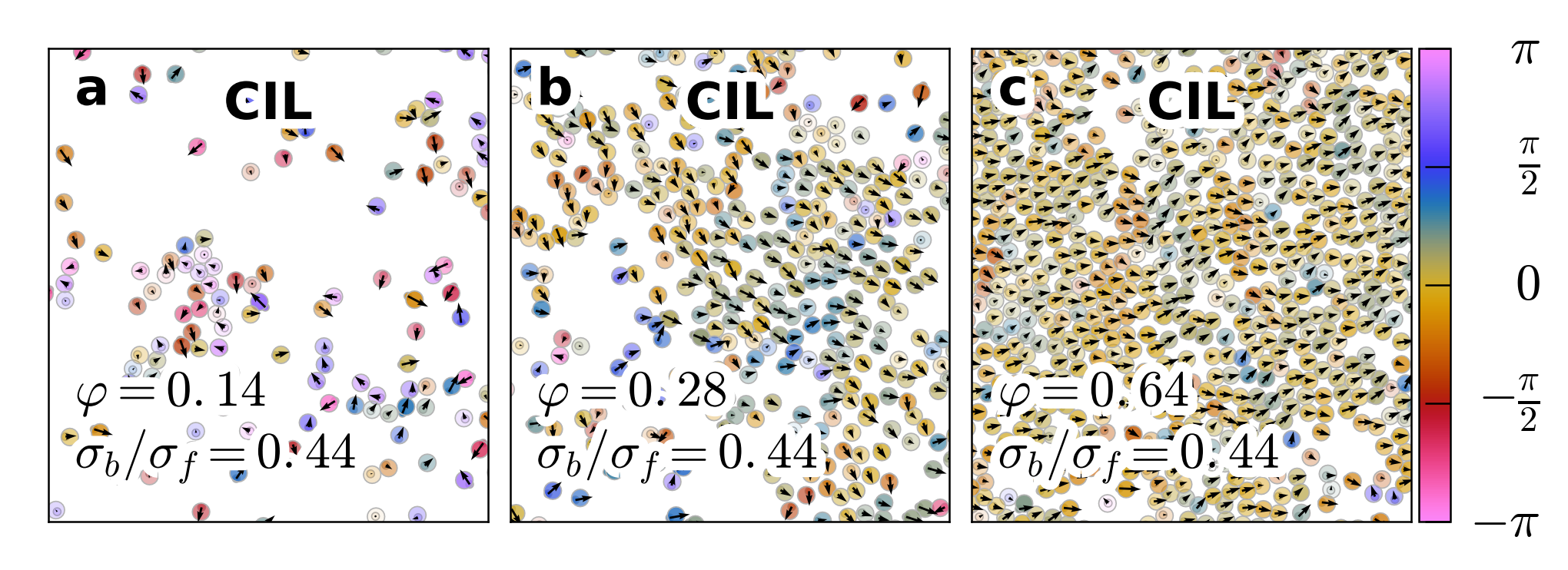}
  \caption{
Simulation snapshots of CIL cells with cell noise in the steady states at different area fractions.
The cell velocities are shown as arrows, the color hue indicates deviation
 from the average directions, and slower cells are lighter in
 color. 
Cells are randomly migrating at a low density (a), some coherency appears
 at an intermediate density (b), and highly coherent migration is observed
 at a high density (c).
Animation videos for these systems are also available, which compare well with similar 
experimental videos for Keratocytes \cite{Szabo2006}.
	}
	\label{fig:with_noise}
\end{figure}

When noise is added, which is the more realistic situation, the order parameter becomes strongly dependent on the area fraction. At small densities, the dynamics are dominated by noise, whereas at large densities collisions between cells play a strong role. This leads to an increase of the order parameter with area fraction in the case of $\sigma_b/\sigma_f = 0.44$ and the order parameter saturates near its noise-free value only at very high densities, see Fig.~3d) of the letter and \cref{fig:with_noise}. This behavior is qualitatively the same as observed in migrating sheets of epithelial cells such as goldfish keratocytes \cite{Szabo2006}.

Genetically identical cells can still have different sizes and structures. This type of cellular noise arising from individual
differences is neglected in the present simulations. It could however be
easily implemented by introducing some polydispersity in
model parameters such as $\sigma_b$, $\sigma_f$, $R_\text{max}$, $m$, and $\kappa$.

\bibliographystyle{apsrev-mod}
\bibliography{supplementalMaterials}